\documentclass[manuscript]{aastex}
\usepackage{natbib}
\usepackage{epsfig}
\slugcomment{}

\begin{document}
\title{Time evolution of flares in GRB 130925A: jet precession in black hole accretion system}
\author{Shu-Jin Hou$^{1,2}$, Tong Liu$^{1,4}$, Wei-Min Gu$^{1}$, Da-Bin Lin$^{3}$, Mou-Yuan Sun$^{1,5}$, Xue-Feng Wu$^{2}$, and Ju-Fu Lu$^{1}$}
\altaffiltext{1}{Department of Astronomy and Institute of Theoretical Physics and Astrophysics, Xiamen University, Xiamen, Fujian 361005, China; tongliu@xmu.edu.cn}
\altaffiltext{2}{Purple Mountain Observatory, Chinese Academy of Sciences, Nanjing 210008, China}
\altaffiltext{3}{Department of Physics and GXU-NAOC Center for Astrophysics and Space Sciences, Guangxi University, Nanning, Guangxi 530004, China}
\altaffiltext{4}{State Key Laboratory of Theoretical Physics, Institute of Theoretical Physics, Chinese Academy of Sciences, Beijing, 100190, China}
\altaffiltext{5}{Department of Astronomy and Astrophysics and Institute for Gravitation and the Cosmos, Pennsylvania State University, University Park, PA 16802, USA}

\begin{abstract}
GRB 130925A, composed of three gamma-ray emission episodes and a series of orderly flares, has been detected by $Swift$, Fermi, Konus-$Wind$, and INTEGRAL. If the third weakest gamma-ray episode can be considered as a giant flare, we find that after the second gamma-ray episode observed by INTEGRAL located at about 2000 s, there exists a positive relation between the time intervals of the adjacent flares and the time since the episode. We suggest that the second gamma-ray episode and its flares originate from the resumption of the accretion process due to the fragments from the collapsar falling back, such a relation may be related to a hyperaccretion disk around a precessed black hole (BH). We propose that the origin and time evolution of the flares, and the approximately symmetrical temporal structure and spectral evolution of the single flare, can be explained well by the jet precession model \citep{Liu2010}. In addition, the mass and spin of the BH can be constrained, which indicates a stellar-mass, fast rotating BH located in the center of GRB 130925A.
\end{abstract}


\section{Introduction}

It is generally believed that the progenitors of short and long gamma-ray bursts (GRBs) are the mergers of two compact objects \citep[see, e.g.,][]{Eichler1989,Paczynski1991} and collapsars of massive stars \citep[see, e.g.,][]{Woosley1993,Paczynski1998}, respectively. In these systems, a rotating black hole (BH) surrounded by a neutrino-dominated accretion flow \citep[NDAF; e.g.,][]{Popham1999,Gu2006,Kawanaka2007,Liu2007,Liu2008,Liu2010,Liu2012,Liu2013,Pan2012,Sun2012,Lei2013,Luo2013,Xue2013} would result, and the fireballs required to power GRBs could be produced via the BZ mechanism \citep{Blandford1977} or annihilation of the neutrino escaping from the NDAFs.

The X-ray flares of GRBs occur after the prompt emission and the time lag is of the order of hundreds or thousands of seconds, which might be related to the activities of the central engine \citep[e.g.,][]{Fan2005,Chincarini2007,Falcone2007,Margutti2011,Luo2013}. Several mechanisms have been proposed to explain the episodic X-ray flares, including gravitational instability in the hyperaccretion disk \citep{Perna2006}, fragmentation of a rapidly rotating core \citep{King2005}, a magnetic switch of the accretion flow \citep{Proga2006}, differential rotation in a post-merger millisecond pulsar \citep{Dai2006}, transition from a thin to a thick disk \citep{Lazzati2008}, He-synthesis-driven wind \citep{Lee2009}, instability in the jet \citep{Lazzati2011}, and outflow caused by the maximal and minimal possible mass accretion rates at each radius of NDAFs \citep{Liu2008}, the episodic jet produced by the magnetohydrodynamic mechanism from the disk \citep{Yuan2012}, and so on. But it is worth noting that the jet precession model \citep[e.g.,][]{Portegies Zwart1999,Reynoso2006,Lei2007,Liu2010,Sun2012,Stone2013} can also explain the origin of some flares. The light curve of GRBs and their flares may be modulated by the jet precession. Most light curves with complex and diverse temporal structures may signal an on-axis/off-axis cycle of the line of sight to a precessed jet axis \citep[e.g.,][]{Portegies Zwart1999,Reynoso2006,Lei2007,Liu2010}.

In a BH accretion system, the misalignment of the angular momenta of two compact objects and anisotropic fall-back material in a collapsar may cause the BH to be precessed by the disk. In this scenario, following the Bardeen-Petterson effect \citep{Bardeen1975}, the inner part of the disk should be aligned with the BH in the accretion process. The different orientation of the angular momentum of the outer disk results in the precessions of the BH and inner disk. Hence, the jet arising from the neutrino annihilation or BZ mechanism above the inner disk is driven to precession \citep[e.g.,][]{Liu2010,Sun2012}. Furthermore, the mass and spin of the BH and the mass of the disk change continuously via the hyperaccretion process, which is related to the precession angle and period, so evolution exists in the precession process.

In this Letter, we focus on samples of the flares in GRB 130925A to test our jet procession model and try to constrain the mass and spin of the central BH. In Section 2, the observations and data analysis are shown and discussed. In Section 3, our jet precession model is introduced and applied to the flares in GRB 130925A. The summary is presented in Section 4.

\section{Observations and data analysis}

GRB 130925A was discovered at 04:11:24 UT on 2013 September 25 by $Swift$/BAT, located at R.A.(J2000) = $+02^h 44^m 42^s.38$, decl.(J2000) =  $-26^\circ 09' 15''.8$ \citep{Lien2013}. The Fermi/GBM was triggered twice and the first pulse may have been a precursor \citep{Fitzpatrick2013}. This burst was also observed by INTEGRAL and Konus-$Wind$ \citep{Savchenko2013,Golenetskii2013}. The source was still detected by BAT two hours after the trigger. The redshift of the burst was $z = 0.347$ \citep{Vreeswijk2013,Sudilovsky2013}.

The $Swift$/XRT begin observing at $147.4 ~\rm s$ after the BAT trigger. The X-ray afterglow of GRB 130925A is very peculiar. It shows several components and is dominated by sharp, bright flares. Figure 1 shows the BAT and XRT light curves, where the data are from the UK Swift Science Data Centre \citep{Evans2009}. The inset shows the light curve with data is from INTEGRAL\footnote{Http://www.apc.univ-paris7.fr/$\sim$savchenk/grb130925a/grb130925a$\_$spiacs.txt.gz}. There are three major gamma-ray emission episodes: the first pulse, which triggered BAT; the second, more fluent episode, from about 1750 s to about 2950 s after the BAT trigger; and the third, the weakest one, the onset of which corresponds to the MAXI detection \citep{Suzuki2013}, from about 3730 s to about 4360 s \citep{Savchenko2013,Golenetskii2013}. The last two episodes were not observed by $Swift$ because the XRT entered Earth-eclipse from 1500 s to 4753 s, but was observed by INTEGRAL and Konus-$Wind$. At the beginning of the X-ray observations, the light curve was a power-law decay with an index of 2.7. After about $750 ~\rm s$, the flares constantly emerged up to about 10000 s, with the peak times of the flares successively at about 1000 s, 1380 s, 4950 s, 7100 s, and 11,200 s \citep{Evans2013}. There is a large gap in the XRT light curve from about 7269 s to about 10,500 s. According to the statistical relation on the mean ratio of the width and peak time of the flares, $\sim 0.13 \pm 0.10$ \citep{Chincarini2007,Margutti2010}, we assume that no flares occurred in the gap. The gamma-ray emission observed by BAT extends to about 1000 s coinciding with the time of the first flare, as well as the emission detected by MAXI at the same time as the weakest episode. After the flares, a power-law decay with index about 0.88 began from about $2\times 10^4 ~\rm s$.

We consider that there are two gamma-ray episodes followed by their own flares. The first gamma-ray episode and its two flares located at about 1000 s and 1380 s may originate from an accretion process. Similarly, including the weakest gamma-ray episode, the second gamma-ray episode at about 2000 s and its flares at about 4950 s, 7100 s, and 11,200 s may arise from another process. If the weakest gamma-ray episode is regarded as a giant flare, we note that time evolution exists for these four flares. Figure 2 shows the positive correlation between the time intervals of the adjacent flares and the time since the second gamma-ray episode in the rest frame with an index of $1.09$. The filled circle on the left bottom of the figure is from the data of the weakest episode at about 4000 s and the flare at about 4950 s, and the other two circles are from the data of the last three flares. The evolution law of the intervals is better than the previous observations of GRB flares \citep[e.g.,][]{Falcone2007,Lazzati2008}, whose characteristics are short-duration and superposed pulses.

Furthermore, we also note that the profiles on the temporal structure and spectral evolution (power-law spectral index $\Gamma$) of the last three flares display approximative symmetry, not the fast-rise-exponential-decay (FRED) phases (see Figure 3), which is very different from previous observations \citep[e.g.,][]{Falcone2007}. Unfortunately, the spectral evolution of the last two flares do not have sufficient confidence because of the inadequate observation data. Besides the time evolution of the flares, the special temporal structure and spectral evolution of the single flare should also be taken into account in models.

In most theoretical models, flares tend to be randomly generated after prompt emission. On the contrary, for the jet precession model, the evolution of the precession periods via the evolution of the BH accretion system is similar to that of the intervals in Figure 2. In this case, if we consider that the intervals of the adjacent flares in the rest frame correspond to the precession periods, the evolution of the intervals can be described, and the characteristics of the BH obtain the limit. Furthermore, the viewing effect in the model can naturally cause these two phases of the light curves. If the jet direction deviates from the line of sight, symmetrical light curves appear.

\section{Jet precession model}

\subsection{The Model}

An accretion disk can be warped by its precession \citep{Sarazin1980}. \citet{Liu2010} and \citet{Sun2012} studied a spinning BH surrounded by a tilted NDAF whose rotation axis is misaligned with that of the BH. The outer part of the disk, whose angular momentum is sufficiently larger than that of the BH, may maintain its orientation and force the BH to precess. Meanwhile, the angular momentum of the inner part of the disk is smaller than that of the BH and should be aligned with the BH spin axis \citep{Bardeen1975}. The ultra-relativistic jet required to power a GRB, is launched by the annihilation of the neutrinos or the BZ mechanism from the inner part of the disk. Its direction is determined by the spin axis of the BH \citep{Popham1999,Liu2007}. When the jet sweeps over the line of sight once, a pulse can be recorded. An on-axis or off-axis cycle of the line of sight to a precessed jet axis corresponds to a FRED or an approximately symmetrical phase in the light curves \citep{Liu2010}.

Using Equations (1), (2) and (5) of \citet{Sun2012} and Equations (5.6) and (5.7) of \citet{Popham1999}, the analytic expression of the precession period $P$ can be written as
\begin{equation}
P = 2793~ a_\ast^{17/13} m^{7/13} \dot{m}^{-30/13} \alpha^{36/13} ~\rm s,
\end{equation}
where $m \equiv M_{\rm BH} / M_\odot$ and $\dot{m} \equiv \dot{M} / (M_\odot ~\rm s^{-1})$, $M_{\rm BH}$  is the mass of the BH, $\dot{M}$ is the accretion rate, $a_\ast$  is the dimensionless spin parameter of the BH, and $\alpha$ is the viscosity parameter of the disk. With $J_{\rm BH}$ representing the angular momentum, the dimensionless of spin parameter is defined as
\begin{equation}
a_\ast = \frac{c J_{\rm BH}}{GM_{\rm BH}^2}.
\end{equation}
It is known that the parameter $a_\ast$ cannot exceed unity, and may increase by the accretion process or decrease by the excess angular momentum of the BH being taken away via the outflow or magnetic field \citep[e.g.,][]{Liu2012,Wu2013}.

Moreover, in the precession system, the mass and spin of the BH should evolve with time during a GRB. The evolution equations of a spinning BH can be written as
\begin{equation}
\frac{d{M}_{\rm BH}}{dt} = \dot{M} e_{\rm ms},
\end{equation}
\begin{equation}
\frac{d{J}_{\rm BH}}{dt} = \dot{M} l_{\rm ms},
\end{equation}
where $e_{\rm ms}$ and $l_{\rm ms}$ are the specific energy and angular momentum at the marginally stable orbit \citep[see e.g.,][]{Liu2012,Wu2013}, which can be written as
\begin{equation}
e_{\rm ms} = \frac{1}{\sqrt{3 \hat{r}_{\rm ms}}} (4- \frac{3 a_\ast}{\sqrt{\hat{r}_{\rm ms}}}),
\end{equation}
\begin{equation}
l_{\rm ms} = 2 \sqrt{3} \frac{G M_{\rm BH}}{c} (1-\frac{2 a_\ast}{3\sqrt{\hat{r}_{\rm ms}}}),
\end{equation}
where $\hat{r}_{\rm ms} = 3+Z_2-[(3-Z_1)(3+Z_1+2Z_2)]^{1/2}$, and $Z_1=1+(1-a_\ast^2)^{1/3}[(1+a_\ast)^{1/3}+(1-a_\ast)^{1/3}]$, $Z_2=(3a_\ast^2+Z_1^2)^{1/2}$. According to the above five equations, the evolution of the BH spin can be expressed as
\begin{eqnarray}
\frac{d{a}_\ast}{dt} =  2\sqrt{3} \frac{\dot{m}}{m} (1-\frac{a_\ast}{\sqrt{\hat{r}_{\rm ms}}})^2 ~{\rm s}^{-1}.
\end{eqnarray}

It is natural to assume that the viscosity parameter is a constant, then the time derivative of $P$ is expressed as
\begin{equation}
\frac{1}{P} \frac{dP}{dt} = \frac{17}{13} \frac{1}{a_\ast} \frac{d a_\ast}{dt} + \frac{7}{13} \frac{\dot{m}}{m} e_{\rm ms} - \frac{30}{13} \frac{1}{\dot{m}} \frac{d \dot{m}}{dt},
\end{equation}
where the unknown viscosity parameter $\alpha$ is eliminated. Besides, the evolution of the precessional angle in the model is based on the transfer of the mass and angular momentum from the disk to the BH.

Furthermore, from Equations (2) to (4), the co-evolution of the mass and spin of the BH can be calculated by
\begin{equation}
m_1=m_0 ~{\rm exp}[\int_{{a_\ast}_0}^{{a_\ast}_1} \frac{4 \sqrt{\hat{r}_{\rm ms}}-3a_\ast}{6(\sqrt{\hat{r}_{\rm ms}}-a_\ast)^2} da_\ast],
\end{equation}
where the subscripts $0$ and $1$ represent the initial and final states, respectively. If we choose ${a_\ast}_0 = 0$ and ${a_\ast}_1=0.998$, $m_1/m_0$ is around $2.2$, and if ${a_\ast}_0 = 0.5$ and ${a_\ast}_1=0.998$, $m_1/m_0$ is around $1.85$.

\subsection{Possible Origin of Flares in GRB 130925A}

We assume that the two gamma-ray episodes and their flares in GRB 130925A originate from different processes. In the collapsar model, we suggest that the first step accretion process powers the first gamma-ray episode and the following two flares, and the resumption of the accretion process due to the fragments falling back \citep[see, e.g.,][]{Kumar2008,Wu2013} powers the second episode and following flares. As we are comparing with gamma-ray episodes, the accretion rate to power flares may be much lower, and combining with Equation (8), the prompt emission cannot be included properly in the time evolution as well as the flares. The first gamma-ray episode and its following flares may also originate from the jet precession process, but only two flares cannot provide information on the evolution of precession. In addition, the two pulses of the weakest gamma-ray episode can be modeled by the precession of the structured jet in a period \citep[e.g.,][]{Portegies Zwart1999,Reynoso2006,Lei2007}, but its unknown information (including X-ray flux, location in afterglow and so on) determine that only the last three flares can be regarded as good samples for testing the model.

There is another noticeable feature of GRB 130925A. Following the BH hyperaccretion, the longer the accretion process is sustained, the wider the deviation of the precessional angle to light of sight is, thus the shapes of the light curves are more symmetrical, just as shown in Figure 3. The cases further verify the possibility that the precession exists in the central engine of GRB 130925A.

The last three flares may be located in the shallow decay phase \citep{Zhang2006} of the afterglow of the second gamma-ray episode. With the observed X-ray luminosity $L_{\rm X,iso} \approx 2.4 \times 10^{48} ~\rm erg~s^{-1}$ at $t_{\rm p} = 5271 ~\rm s$ and $L_{\rm X,iso}=\eta \dot{m} M_\odot c^2$, where $\eta=0.01$ including the efficiency and beaming effect, the mass accretion rate can be estimated as $\dot{m}_{\rm p}\approx0.0002$, which satisfies the requirement of the disk mass, about several solar mass \citep[e.g.,][]{Popham1999}. The time evolution of the accretion rate can be expressed as
\begin{equation}
\dot{m}=\dot{m}_{\rm p} (\frac{t-t_0}{t_{\rm p}-t_0})^{-0.5},
\end{equation}
where $t_0$ is the exploded time of the first flare, which is around $742 ~\rm s$. The weakest gamma-ray episode corresponds to a larger mass accretion rate than those of the last three flares, which does not comply with the power-law index $\sim 0.5$ in Equation (10). The reason that the filled circle on the bottom left of Figure 2 deviates from the fitting line of the last two circles may be related to this situation.

The time intervals of the adjacent three flares in the rest frame correspond to the precession periods, which are about $1596~\rm s$ and $3044~\rm s$, so the mean period is about $2320~\rm s$. The change rate of the time interval in the rest frame corresponds to that of the period $dP/dt$, which is about $0.62$. According to these two data points and Equation (8), we can plot the relation of the mass and spin of the BH in the final state as shown by the solid line in Figure 4.

Here two restrictive rules are given. The first is the co-evolution of the mass and spin of the BH corresponding to Equation (9). We reasonably assumed that $a_{\ast 0}$ is $0.8$ for an original spinning BH after a hyperaccretion process lasting hundreds of seconds, and approximatively give $(m_1-m_0)$ by $\Delta m = \int_{t_0}^{t_{\rm p}} \dot{m} dt$ if the relativistic factor $e_{\rm ms}$ is ignored, then $a_{\ast 1}$ can be calculated. The relation between $a_{\ast 1}$ and $m_1$ is shown by the dashed line in Figure 4. Moreover, as the second rule, in the collapsar model, it is reasonable that $m_0$ should be larger than $3$. Also, after a long-duration accretion process, we give the second limit using $m_1=m_0+\Delta m$, which is shown by the dotted line in Figure 4. Thus the thick solid line is located in an upper right region to the dashed and dotted lines, which means that these conclusions stand these two imperative tests. Furthermore, we consider that the final BH spin is extremely high since the accretion has lasted nearly 6000 s. Consequently, the BH mass in the final state may be about $10 ~M_\odot$, which is also consistent with the collapsar models \citep{Popham1999}.

In addition, we choose two typical points in the thick solid line of Figure 4, (15.20, 0.95) and (9.04, 0.99), which correspond to the viscosity parameter $\alpha = 4.66\times 10^{-4}$ and $5.04 \times 10^{-4}$, respectively, as well as the value of $\alpha$ around $5 \times 10^{-4}$ calculated by other data using Equation (1). Such a low viscosity is reasonable because low $\alpha$ is required by a long-duration accretion process with low accretion rate in NDAF model (the viscous timescale corresponding to the duration of the burst is inversely proportional to the viscosity parameter). From an energetic perspective, \citet{Chevalier1996} suggested that the value of $\alpha$ should be small in order to allow the balance between viscous heating and neutrino cooling in the neutrino-cooled accretion disk.

\section{Summary}

In this Letter, we have proposed that the jet precession model \citep{Liu2010} can explain the origin and time evolution of the flares, and the approximately symmetrical temporal structure and spectral evolution of the last three flares in GRB 130925A. As a consequence, the mass and spin of the BH can be constrained by the observation, which indicates that a stellar-mass, fast rotating BH may exist in the center of GRB 130925A and it is precessed by a massive accretion disk. Recently, we noticed that there might also exist evidence of precession in the giant X-ray bump of GRB 121027A.

For the BH hyperaccretion model, there are two mechanisms to power GRBs. One is the neutrino radiation from the disk and annihilation above the disk, and the other is the extraction of rotational energy of the BH by the magnetic field, such as the BZ mechanism and the magnetic coupling effect \citep[e.g.,][]{Lei2013,Luo2013}. Meanwhile, the magnetic field is a possible medium to transfer the angular momentum of the BH to outer space, and the rigorous accretion rates are not essential to BZ mechanism. The jet precession in the magnetized NDAF model is worth studying.

\acknowledgments
We thank Wei-Hua Lei and Ya-Ping Li for beneficial discussions and the anonymous referee for very useful suggestions and comments. This work made use of data supplied by the UK Swift Science Data Centre at the University of Leicester. This work is supported by the National Basic Research Program of China (973 Program) under grants 2013CB834900 and 2014CB845800, and the National Natural Science Foundation of China under grants 11103015, 11222328, 11233006, 11322328, 11333004, and U1331101. X. F. Wu acknowledges support by the One-Hundred-Talents Program and the Youth Innovation Promotion Association of Chinese Academy of Sciences. M. Y. Sun acknowledges support by the China Scholarship Council (No. [2013]3009).

\clearpage

\begin{figure}
\centering
\includegraphics[angle=0,scale=0.6]{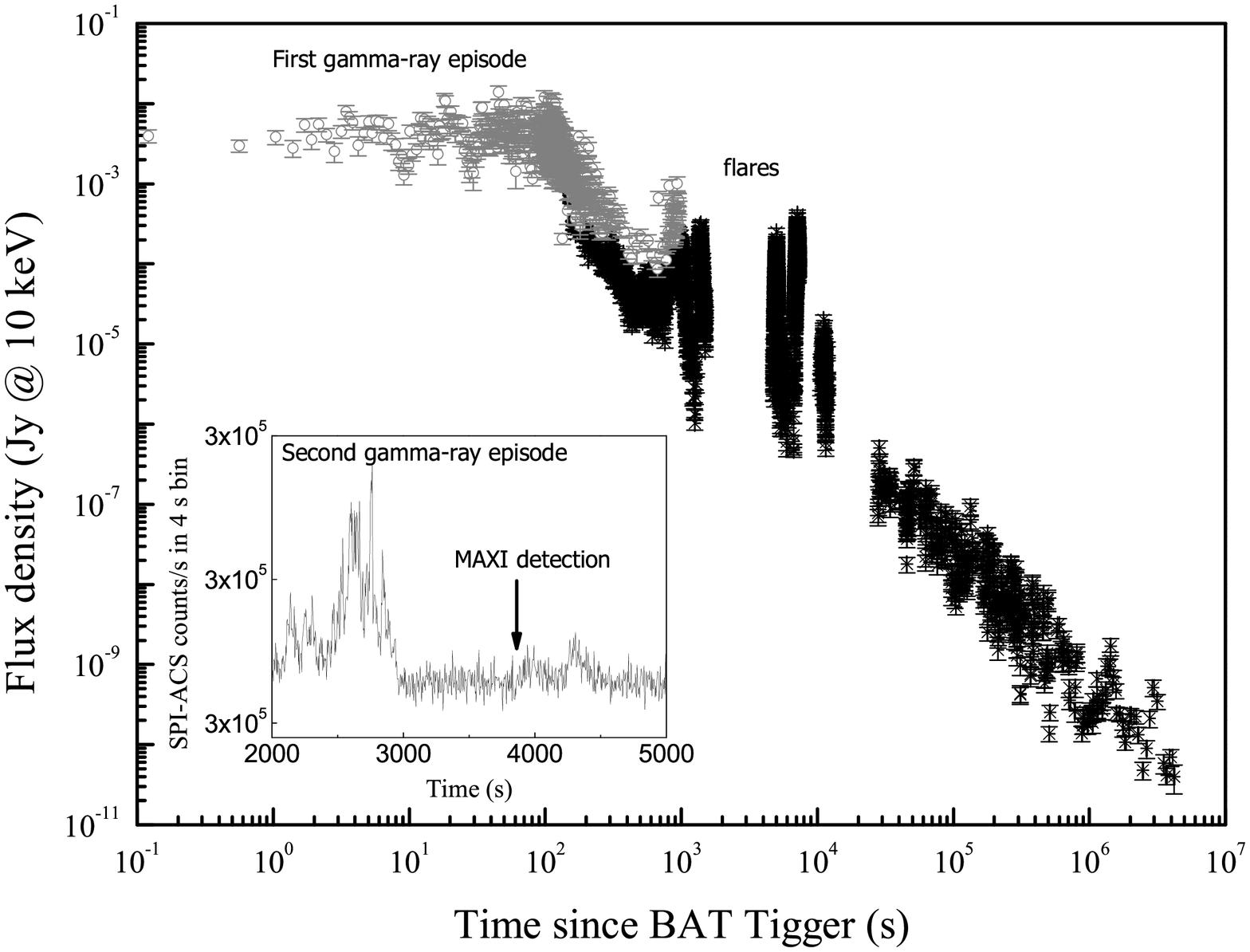}
\caption{The BAT (gray) and XRT (black) light curves of GRB 130925A. The inset shows the light curve observed by INTEGRAL.}
\end{figure}

\clearpage

\begin{figure}
\centering
\includegraphics[angle=0,scale=0.6]{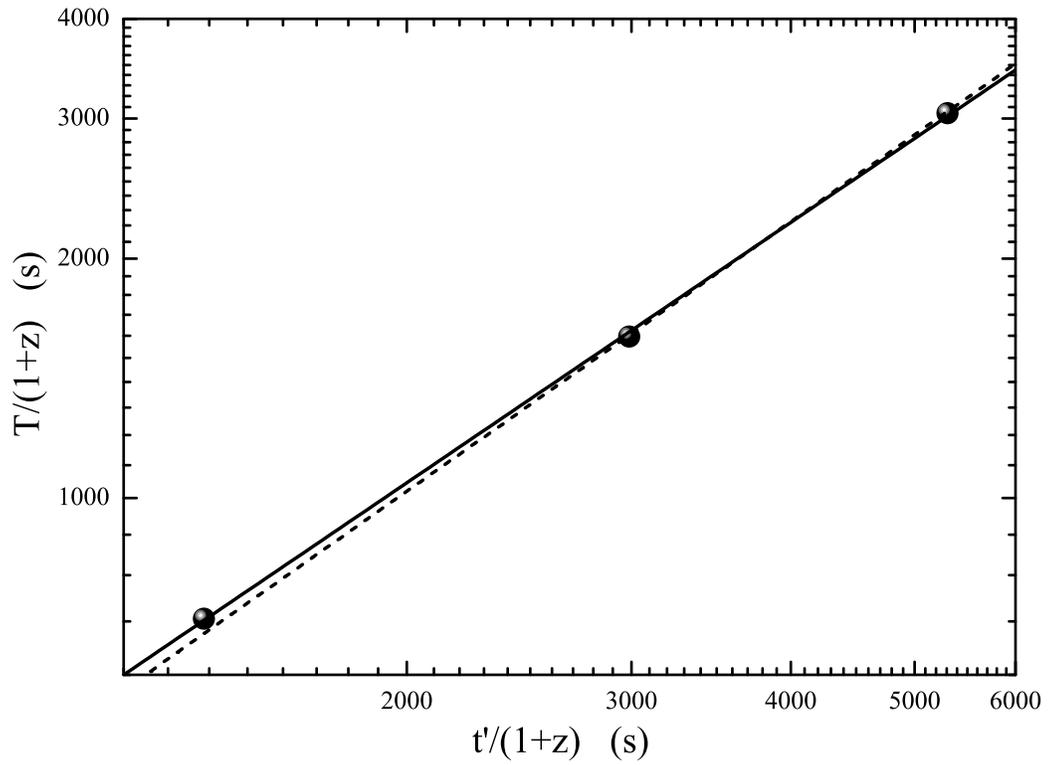}
\caption{Evolution of the time intervals $T$ between adjacent flares after the second gamma-ray episode of GRB 130925A with time since the episode $t'$ in the rest frame. The solid and dashed lines represent the fitting lines of all the three filled circles and the last two circles with the power-law indices of 1.09 and 1.12, respectively.}
\end{figure}

\clearpage

\begin{figure}
\centering
\includegraphics[angle=0,scale=0.35]{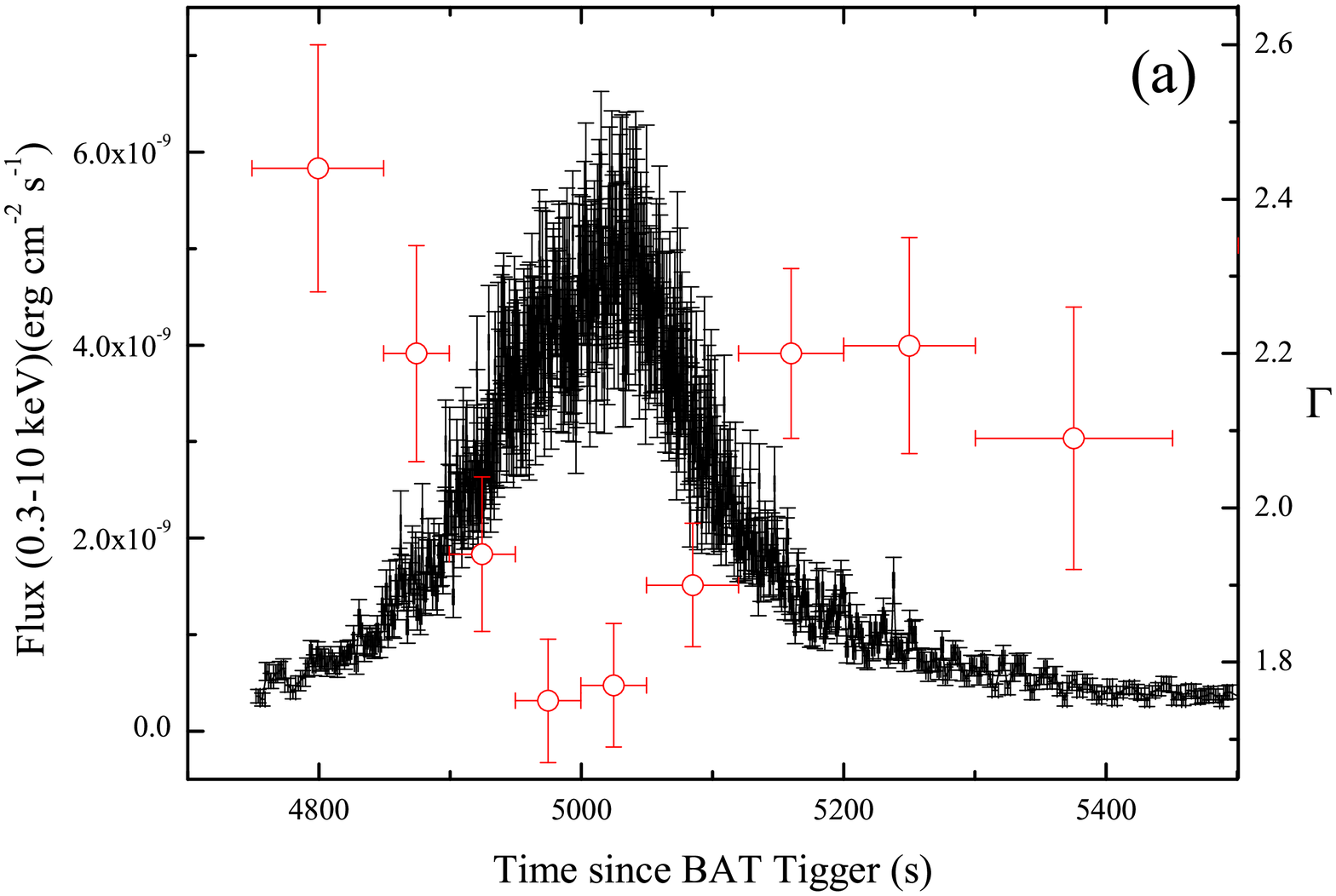}
\includegraphics[angle=0,scale=0.35]{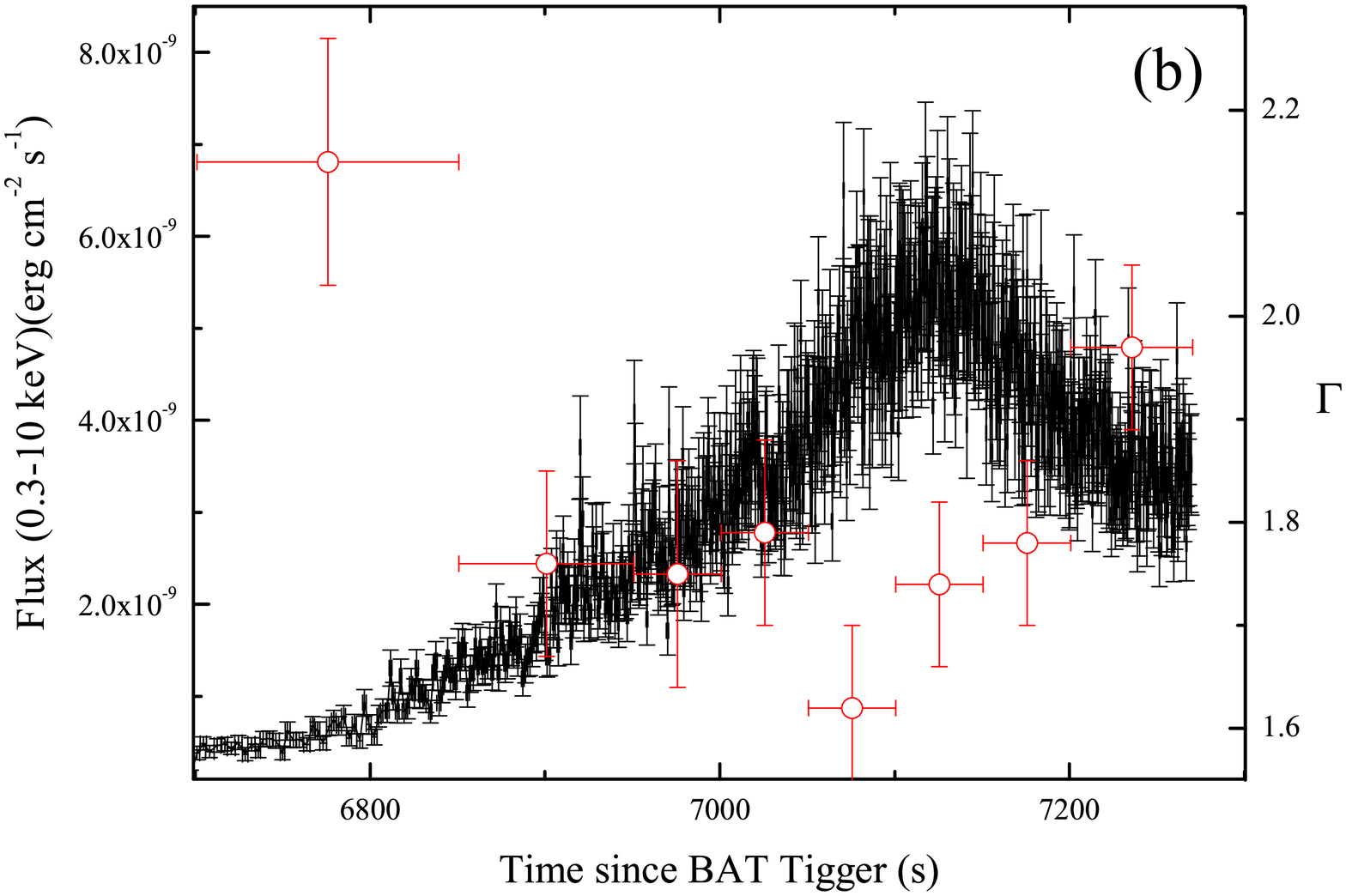}
\includegraphics[angle=0,scale=0.35]{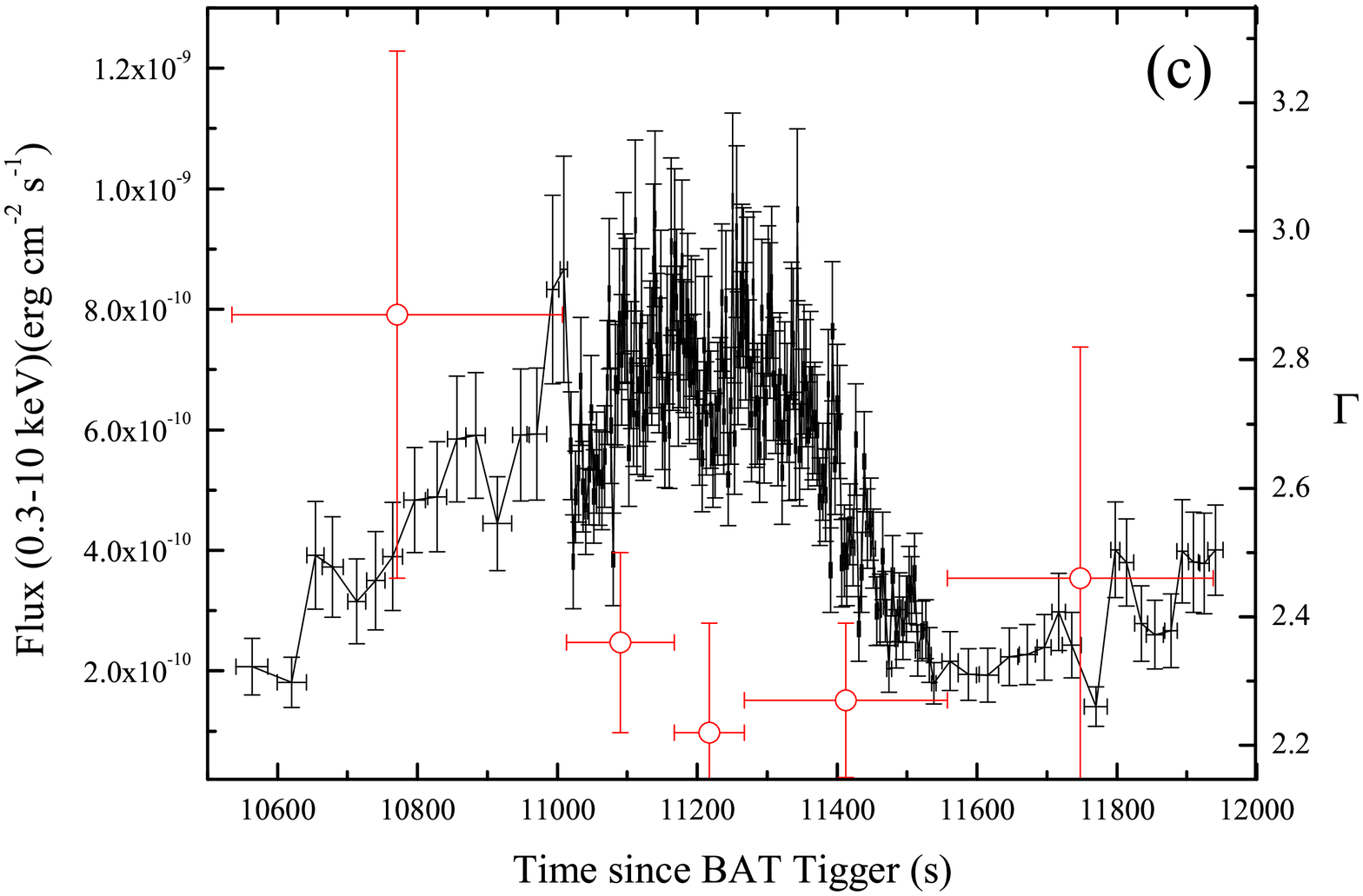}
\caption{Temporal structure and spectral evolution (power-law spectral index $\Gamma$) of the last three flares. The values of $\Gamma$ are represented by the red open circles.}
\end{figure}

\clearpage

\begin{figure}
\centering
\includegraphics[angle=0,scale=0.6]{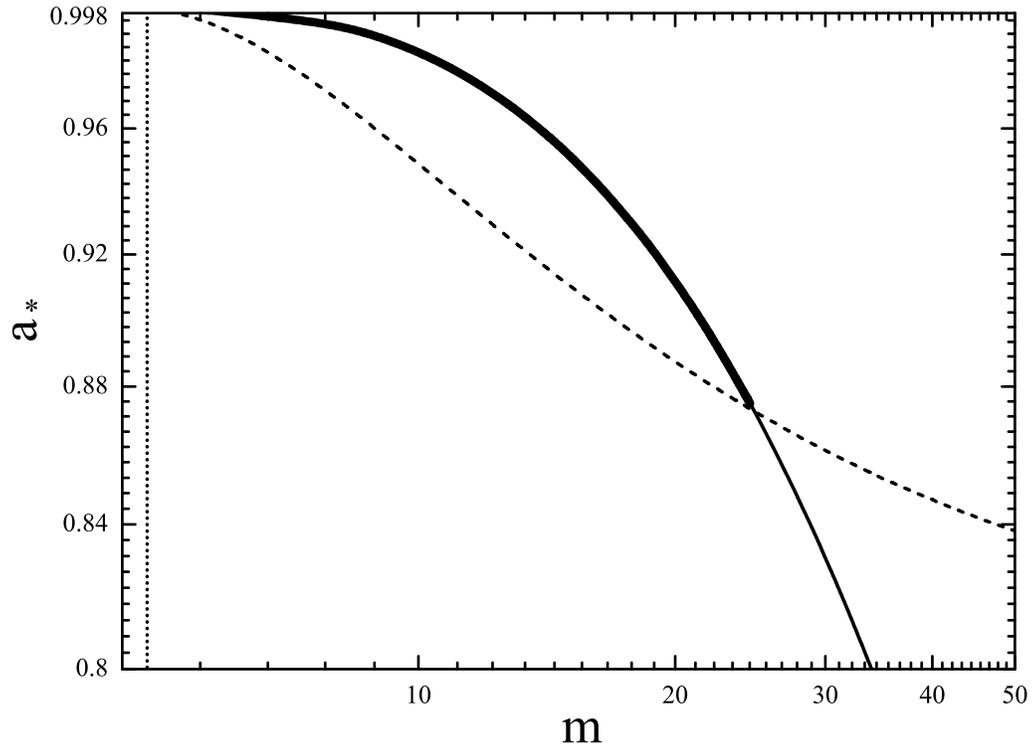}
\caption{Possible range of the mass and spin of the central BH (thick solid line). The dotted line represents the lower limit of the BH mass estimated by the accretion rate and period, and the dashed line shows the constraint on the mass and spin calculated by Equation~(9).}
\end{figure}

\clearpage

\end{document}